\documentclass[12pt]{article}
\usepackage{amsmath}
\usepackage{amssymb}
\usepackage{amsfonts}

\newtheorem{theorem}{Theorem}
\newtheorem{acknowledgement}[theorem]{Acknowledgement}

\input{tcilatex}
\begin{document}

\title{Guarantees in Fair Division: general or monotone preferences}
\author{Anna Bogomolnaia$^{\spadesuit }$ and Herv\'{e} Moulin$^{\spadesuit }$%
}
\date{September 2020\\
$\spadesuit $ University of Glasgow, UK, and Higher School of Economics, St
Petersburg, Russia}
\maketitle

\begin{abstract}
To divide a "manna" $\Omega $ of private items (commodities, workloads,
land, time intervals) between $n$ agents, the worst case measure of fairness
is the welfare guaranteed to each agent, irrespective of others'
preferences. If the manna is non atomic and utilities are continuous (not
necessarily monotone or convex), we can guarantee the minMax utility: that
of our agent's best share in her worst partition of the manna; and implement
it by Kuhn's generalisation of Divide and Choose. The larger Maxmin utility
-- of her worst share in her best partition -- cannot be guaranteed, even
for two agents.

If for all agents more manna is better than less (or less is better than
more), our Bid \& Choose rules implement guarantees between minMax and
Maxmin by letting agents bid for the smallest (or largest) size of a share
they find acceptable.

\begin{acknowledgement}
We are grateful for the critical comments of Haris Aziz, Steve Brams, Eric
Budish, Ariel Procaccia, Richard Stong, and participants in seminars and
workshops at LUISS University, UNSW Sydney, Universities of Gothenburg, St
Andrews, the London School of Economics, the 12th International Symposium on
Algorithmic Game Theory, Athens, and webinars at at ETH Zurich and Caltech.
Special thanks to Erel Segal-Halevi and Ron Hozman for their detailed
comments on the first version of this paper.

Bogomolnaia and Moulin acknowledge the support from the Basic Research
Program of the National Research University Higher School of Economics.
\end{acknowledgement}
\end{abstract}

\section{Introduction and the punchlines}

The fair division of a common property manna -- resources privately consumed
-- \ is a complicated problem if its joint owners have heterogenous
preferences over the manna. A coarse yet important benchmark is the welfare 
\textit{Guarantee} a division rule offers to each participant: this is the
highest welfare that a given agent can secure in this rule, irrespective of
the preferences of other agents, even if our agent is clueless about the
latter and assumes the worst. The more an agent is risk averse and the less
she knows about others' preferences, the more this worst case benchmark
matters to her.

Our goal is to throw some light on the feasible Guarantees in the very
general class of \textit{non atomic }fair division problems: small changes
in the size of a share result in small utility changes (a continuity
property explained below). Our model places no other restrictions on the
structure of preferences and corresponding utilities, or their direction:
the manna may contain some desirable parts (money, tasty cake, valuable
commodities), some not (unpleasant tasks, financial liabilities, burnt parts
of the cake that must still be eaten \cite{SHa}); agents may disagree over
which parts are good or bad; utilities can be single-peaked over some parts
(teaching loads, volunteering time, shares of a risky project),
single-dipped on others, etc..

Assume that the manna $\Omega $ and the domain $\mathcal{D}$ of potential
preferences are common knowledge, and define a \textit{Fair Guarantee} as a
mapping $(u_{i},n)\rightarrow \Gamma (u_{i};n)$ selecting for each
preference in $\mathcal{D}$, described for clarity as a utility function $%
u_{i}$, and each number $n$ of joint owners, a utility level. The mapping is
fair because it ignores agent $i$'s identity, and it must be feasible: for
any profile $(u_{i})_{i=1}^{n}$ of utilities in $\mathcal{D}^{n}$ there
exists a partition $(S_{i})_{i=1}^{n}$ of $\Omega $ such that $%
u_{i}(S_{i})\geq \Gamma (u_{i};n)$ for all $i$.

Given the division problem $(\Omega ,\mathcal{D})$ we ask what are the best
(highest) Fair Guarantees? and what mechanism implements\footnote{%
In the simple sense of implementation described in the last paragraph of
this section.} them?

Observe first that any Fair Guarantee $\Gamma (u;n)$ is bounded above by the
utility, denoted $Maxmin(u;n)$, of the worst share for $u$ in the best $n$%
-partition of the manna. For all $u\in \mathcal{D}$ and $n$ we have%
\begin{equation}
\Gamma (u;n)\leq Maxmin(u;n)=\max_{\Pi =(S_{i})_{i=1}^{n}}\min_{1\leq i\leq
n}u(S_{i})  \label{14}
\end{equation}%
where the maximum (that may not be achieved exactly) bears on all $n$%
-partitions $\Pi =(S_{i})_{i=1}^{n}$ of $\Omega $. This follows by
feasibility of $\Gamma (u;n)$: at the unanimous profile where $u_{i}=u$ for
all $i$ there is a partition $\Pi $ such that $u(S_{i})\geq \Gamma (u;n)$
for all $i$, hence $\Gamma (u;n)\leq \min_{1\leq i\leq n}u(S_{i})\leq
Maxmin(u;n)$.

Therefore if $(u,n)\rightarrow Maxmin(u;n)$ is itself a Fair Guarantee (it
is fair, but the issue is feasibility), it is the best possible one and
answers the first of the two general questions above. This happens in two
well known and much discussed families of fair division problems.

In the cake-cutting model due to Steinhaus (\cite{St}) the manna $\Omega $
is a measurable space endowed with a non atomic measure, and utilities are
additive measures, absolutely continuous with respect to the base measure.
Additivity of $u$ implies $Maxmin(u;n)\leq \frac{1}{n}u(\Omega )$; this is
in fact an equality because the cake can be partitioned in $n$ shares of
equal utility. Agent $i$'s share $S_{i}$ is \textit{Proportionally Fair }if $%
u_{i}(S_{i})\geq \frac{1}{n}u_{i}(\Omega )$: this is feasible for all agents
at any preference profile $(u_{i})_{i=1}^{n}$, therefore Proportional
Fairness offers the best possible Guarantee in this model, and is the
weakest and least controversial test of fairness throughout the cake-cutting
literature (\cite{BT2} and \cite{RW}).

In the microeconomic model of fair division the manna is a bundle $\omega
\in 
\mathbb{R}
_{+}^{K}$ of $K$ divisible and non disposable items, and $\mathcal{D}$ is
the set of convex and continuous preferences over $[0,\omega ]$ (not
necessarily monotonic). It is feasible to give an equal share $\frac{1}{n}%
\omega $ to every agent, so that $\Gamma ^{es}(u;n)=u(\frac{1}{n}\omega )$
is a feasible Guarantee. In $\mathcal{D}$ the inequality $Maxmin(u;n)\leq u(%
\frac{1}{n}\omega )$ is also true.\footnote{%
Pick a hyperplane $H$ supporting the upper contour of $u$ at $\frac{1}{n}%
\omega $; the \textit{lower} contour of $u$ at $\frac{1}{n}\omega $ contains
one closed half-space cut by $H$, and every division of $\omega $ as $\omega
=\sum_{1}^{n}z_{i}$ includes at least one $z_{j}$ in that half-space.}
Therefore the optimal Guarantee is $\Gamma ^{es}$, aka the \textit{Equal
Split lower bound} $u_{i}(z_{i})\geq u(\frac{1}{n}\omega )$ (where $z_{i}$
is $i$'s share of $\omega $). Here too it is the starting point of the
discussion of fairness (see e. g., \cite{Th} and \cite{Mo3}).\smallskip

As soon as we drop either additivity in the former model, or convexity in
the latter one, the Maxmin benchmark is not a Fair Guarantee any more:
already in some two person problems no division of the manna yields at least 
$Maxmin(u_{i};2)$ for both $i=1,2$. In a simple example Ann and Bob share $%
10 $ units of a single non disposable divisible item (e.g., time spent in a
given activity). Ann's preferences are single-peaked (hence convex), while
Bob's are single-dipped (\textit{see Figure 1}):%
\begin{equation*}
u_{A}(x)=x(12-x)\text{ \ ; \ }u_{B}(x)=x(x-6)\text{ for }0\leq x\leq 10
\end{equation*}%
Compute%
\begin{equation*}
Maxmin(u_{A})=35\text{\ at }\Pi _{1}=\{5,5\}\text{ ; }Maxmin(u_{B})=0\text{\
at }\Pi _{2}=\{0,10\}
\end{equation*}%
If Bob's share is worth at least $Maxmin(u_{B})$ then Ann gets either the
whole manna or at most $4$ units: so her utility is at most $32$ and we see
that $(Maxmin(u_{A}),Maxmin(u_{B}))$ is not feasible.

A second critical benchmark utility is $minMax(u;n)$, the utility of the
best share for $u$ in the worst possible $n$-partition of $\Omega $:%
\begin{equation*}
minMax(u;n)=\min_{\Pi =(S_{i})_{i=1}^{n}}\max_{1\leq i\leq n}u(S_{i})
\end{equation*}%
where as before the minimum bears on all $n$-partitions of $\Omega $.

Our first main result, Theorem 1 in Section 4, says that in any non atomic
problem, the mapping $u\rightarrow minMax(u;n)$ is a Fair Guarantee; in
particular $minMax(u;n)\leq Maxmin(u;n)$ for all $u\in \mathcal{D}$ and $n$
(by (\ref{14})). Moreover the $minMax$ Guarantee is implemented by Kuhn's
little known $n$-person generalisation of Divide and Choose (\cite{Kuh}),
denoted here D\&C$_{n}$.

The result is clear in two person problems, where ordinary Divide and Choose
clearly guarantees her $Maxmin$ to the Divider and his $Minmax$ to the
Chooser. For instance in the example above Ann would Divide as $\Pi
_{1}=\{5,5\}$ and Bob would get utility $-5$, exactly his $minMax(u_{B};2)$;
while Bob would Divide as $\Pi _{2}=\{0,10\}$, and Ann would Choose $10$,
thus achieving $minMax(u_{A};2)=20$.

In three persons problems D\&C$_{3}$ works as follows. The Divider Ann
offers a $3$-partition $\Pi =\{S_{1},S_{2},S_{3}\}$ where all shares are of
equal value to her; Bob \textit{accepts} \textit{all shares }worth at least $%
minMax(u_{B};3)$, and Charles all those worth least $minMax(u_{C};3)$. If
Bob and Charles can each be assigned a share they accept, we do so and Ann
gets the last piece.\footnote{%
Maybe more than one such assignment is feasible; any choice implements the
target Guarantee, which is all we need.} If both accept a single share in $%
\Pi $, the same one, we give one of the remaining shares $S_{k}$ to Ann (it
does not matter which one) and then run D\&C$_{2}$ between Bob and Charles
for $\Omega \diagdown S_{k}$ (it does not matter who Divides or Chooses).

The $n$-person division rule D\&C$_{n}$ proceeds similarly in at most $n-1$
steps of Division and Acceptance between a shrinking set of agents sharing a
shrinking manna. Its only subtlety is a simple combinatorial matching step
(Lemma 2 in Section 4) after each partitioning of the remaining manna.

The hard step in proving Theorem 1 is Lemma 1 in Subsection 3.2, stating
that in each round of D\&C$_{n}$ the current Divider can find an \textit{%
equipartition}: a partition of the remaining manna where all shares are
equally valuable to this Divider. Because we only assume that he manna is
measurable and endowed with a non atomic measure, and that utilities are
continuous in that measure, the proof of Lemma 1 requires advanced tools in
algebraic geometry: this the object of the companion paper \cite{AvKa2}, see
the discussion in Subsection 3.2.\medskip 

Our second main result, Theorem 2 in Subsection 5.2, focuses on non atomic
problems where preferences are also \textit{co-monotone}: that is, \textit{%
increasing} if enlarging a share cannot make it worse and we speak of a 
\textit{good manna}; or \textit{decreasing} if the opposite holds and we
have a \textit{bad manna}. Either restriction on preferences opens the door
to a new family of division rules significantly simpler than D\&C$_{n}$ and
implementing a higher Guarantee than the $minMax$. These rules are inspired
by the well known family of \textit{Moving Knife} (MK$_{n}$) rules (Dubins
and Spanier \cite{DS}) that we recall first.

Assume the manna is good: a knife cuts continuously an increasing share of
the cake; agents can stop the knife at any time; the \textit{first} agent
who does gets the share cut so far. Repeat between the remaining agents and
manna. For a bad manna, agents can drop\ at any time and the \textit{last }%
one to drop gets the share cut so far.

A Moving Knife (MK) rule chooses a single arbitrary path for the knife,
which tightly restricts the range of individual shares and partitions, hence
can result in a very inefficient allocation. We introduce a large family of
rules in the same spirit as MK but with all partitions in their range, that
we call the \textit{Bid \& Choose} (B\&C$_{n}$) rules. Each rule is defined
by fixing a benchmark \textit{additive measure} of the shares, diversely
interpreted as their size, their market price, etc.. If the manna is good a
bid $b_{i}$ by agent $i$ is the smallest measure of a share that $i$ finds
acceptable: the smallest bidder $i^{\ast }$ chooses freely a share of
measure at most $b_{i^{\ast }}$, then we repeat between the remaining agents
and manna. For a bad manna the bid $b_{i}$ is the largest size of a share
that $i$ finds acceptable, and the largest bidder $i^{\ast }$ picks any
share of size at least $b_{i^{\ast }}$.

Theorem 2 in Section 5 shows that all B\&C$_{n}$ rules, as well as all MK$%
_{n}$ rules implement a Guarantee between the $minMax$ and $Maxmin$ level.

A handful of examples in Subsection 5.3 show that the B\&C$_{n}$ Guarantee
improves substantially the $minMax$ Guarantee in the microeconomic model of
fair division. There the Equal Split Guarantee is the $Maxmin$ benchmark
(the best possible) for agents with convex preferences, while for agents
with \textquotedblleft concave\textquotedblright\ preferences (convex lower
contours) Equal Split is the $minMax$ Guarantee, which the B\&C$_{n}$
Guarantee improves significantly.\smallskip 

Throughout the paper we speak of implementation in the very simple sense
adopted by most of the cake cutting literature (e. g., \cite{BT2}), and
formalized in the general collective decision context as implementation in
\textquotedblleft protective equilibrium\textquotedblright\ (Barbera and
Dutta \cite{BaDu}). A rule implements (guarantees) a certain utility level $%
\gamma $ means this: no matter what her preferences, each agent has a
strategy that depends also upon $\Omega ,n$ and $\mathcal{D}$, such that
whatever other agents do the utility of her share is no less than $\gamma $.
Moreover the \textquotedblleft guaranteeing strategy\textquotedblright\ is
essentially unique.

\section{Relevant literature}

The two welfare levels $Maxmin$ and $minMax$ are key to our results. In the 
\textit{atomic }model where the manna is a set of indivisible items, they
are introduced by Budish (\cite{Bud}) and Bouveret and Lemaitre (\cite{BoLe}%
) respectively . If utilities are additive in that model, the basic
inequality of our \textit{non atomic} model is reversed:%
\begin{equation*}
Maxmin(u;n)\leq \frac{1}{n}u(\Omega )\leq minMax(u;n)
\end{equation*}%
and $minMax(u;n)$ is obviously not a feasible Guarantee. It took a couple of
years and many brain cells to check that the $Maxmin$ lower bound may not be
feasible either for three or more agents (\cite{PW}), though this happens in
rare instances of the model (\cite{KPW}).\footnote{%
If the manna is atomic and utilities are not necessarily additive, it is
easy to construct examples showing that all six orderings of $Maxmin$, $%
minMax$, and $\frac{1}{n}u(\Omega )$ are possible.} Our paper is the first
systematic discussion of these two bounds in the non atomic model of cake
division.

Kuhn's 1967 $n$ person generalisation of Divide and Choose (\cite{Kuh})
promptly implements the $minMax$ guarantee in our model: Theorem 1. Except
for a recent discussion in \cite{AHSeg} for additive utilities, D\&C$_{n}$
has not received much attention, a situation which our paper may help to
correct. In particular, unlike the Diminishing Share (\cite{St}) Moving
Knife (\cite{DS}), and Bid and Choose rules, it is very well suited to
divide \textit{mixed manna, }i. e., containing subjectively good and bad
parts, as when we divide the assets and liabilities of a dissolving
partnership. Introduced in \cite{BMSY1} and \cite{BMSY2} for the competitive
fair division of commodities in the microeconomic model, the mixed manna
model is discussed by \cite{SHa} for a general cake, and by \cite{ACIW} for
indivisible items.

Privacy preservation is a growing concern in a world of ever expanding
information flows. The D\&C$_{n}$ rule stands out for its informational
parsimony: each Divider only reports a partition with the understanding that
she is indifferent between the two shares she just cut, and Choosers only
only accept a subset of these shares. If the manna is mixed, no one is asked
to explain which parts they view as good or bad: for instance if we divide
tasks, I may not want others to know which tasks I am actually happy to
perform, or which ones are very painful to me.

The \textquotedblleft cuts\textquotedblright\ selected\ by Dividers and
\textquotedblleft queries\textquotedblright\ answered by Choosers require
only a modest cognitive effort: no one needs to form complete preference
relations over all shares of the cake. Taking this feature to heart, a large
literature in the cake cutting model evaluates the informational complexity
of various mechanisms by the number of \textquotedblleft
cuts\textquotedblright\ and \textquotedblleft queries\textquotedblright\
they involve: see \cite{BT2} or \cite{RW}, and more recently \cite{CF} and 
\cite{CNS}. This line of research goes beyond the test of Proportional
Guarantee, and find cuts and queries division rules more complex than D\&C$%
_{n}$ reaching an Envy-free division of the cake. The algorithms in Brams
and Taylor (\cite{BT1}), and more recently Aziz and McKenzie (\cite{AMK}),
do exactly this when utilities are additive and non atomic; but because they
involve an astronomical number of cuts and queries they are of no practical
interest and squarely contradict informational parsimony. See (\cite{Bran}, 
\cite{KLP}) for some fine tuning of these general facts.

The \textquotedblleft equipartition\textquotedblright\ Lemma (Subsection
3.2) is critical to the proof of Theorem 1, and proved in \cite{AvKa2} by
algebraic geometry techniques. These, or subtle variants of Sperner's Lemma,
demonstrate the existence of an Envy-free division under very general
preferences, where which share I like best in a given partition can depend
upon the partition itself, not just upon my own share: Stromquist's (\cite%
{Str}) and Woodall's (\cite{Woo}) seminal insights are considerably
strenghtened by the recent results in \cite{Su}, \cite{SHa}, \cite{MeZe} and 
\cite{AvKa}. However all these results assume that, either all agents
(weakly) prefer any non empty share to the empty share, or all weakly prefer
the empty share to any non empty one: this rules out a mixed manna.

We noted earlier that the concept of \textit{unanimity utility} (the common
efficient utility level in the economy where everyone has the same
preferences) leads to the Equal Split Guarantee when we divide private goods
and preferences are convex (see Footnote 2). When applied to fair division
problems involving production, it defines some compelling Fair Guarantees as
well as some meaningful \textit{upper bounds} on individual welfare: \cite%
{Mo2}, \cite{Mo1}.

\section{Non atomic fair division}

\subsection{Basic definitions}

The manna $\Omega $ is a bounded measurable set in an euclidian space,
endowed with the Lebesgue measure $|\cdot |$, and such that $|\Omega |>0$. A
share $S$ is a possibly empty measurable subset of $\Omega $, and $\mathcal{B%
}$ is the set of all shares. A $n$-partition of $\Omega $ is a $n$-tuple of
shares $\Pi =(S_{i})_{i=1}^{n}$ such that ${\large \cup }_{i=1}^{n}S_{i}=%
\Omega $ and $|S_{i}\cap S_{j}|=0$ for all $i\neq j$; and $\mathcal{P}%
_{n}(\Omega )$ is the set of all partitions of $\Omega $. We define
similarly an $n$\textit{-partition of }$S$ for any share $S\in \mathcal{B}$,
and write their set as $\mathcal{P}_{n}(S)$.

If $S\otimes T=(S\cup T)\diagdown (S\cap T)$ is \ the symmetric difference
of shares, recall that $\delta (S,T)=|S\otimes T|$ is a pseudo-metric on $%
\mathcal{B}$ (a metric except that $\delta (S,T)=0$ iff $S$ and $T$ differ
by a set of measure zero).

A utility function $u$ is a mapping from $\mathcal{B}$ into $%
\mathbb{R}
$ such that $u(\varnothing )=0$ and $u$ is continuous for the pseudo-metric $%
\delta $ and bounded. So $u$ does not distinguish between two shares at
pseudo-distance zero (equal up to a set of measure zero): for instance $%
u(S)=0$ if $|S|=0$. Also if the sequence $|S^{t}|$ converges to zero in $t$,
so does $u(S^{t})$. We write $\mathcal{D}(\Omega )$ for this domain of
utility functions.

So a non atomic division problem consists of $(\Omega ,\mathcal{B}%
,(u_{i})_{i=1}^{n}\in \mathcal{D}(\Omega )^{n})$.

Several subdomains of $\mathcal{D}(\Omega )$ play a role below:

\begin{itemize}
\item additive utilities: $u\in \mathcal{A}dd(\Omega )$ \textit{iff} $%
u(S)=\int_{S}f(x)dx$ for all $S$, where $f$ is bounded and measurable in $%
\Omega $;

\item monotone increasing: $u\in \mathcal{M}^{+}(\Omega )$ \textit{iff} $%
S\subset T\Longrightarrow u(S)\leq u(T)$ for all $S,T$;

\item monotone decreasing: $u\in \mathcal{M}^{-}(\Omega )$ \textit{iff} $%
S\subset T\Longrightarrow u(S)\geq u(T)$ for all $S,T$;

\item separable: $u\in \mathcal{S}(\Omega )$ \textit{iff} there is a finite
set $A$, a partition $(C_{a})_{a\in A}\in \mathcal{P}_{|A|}(\Omega )$ of $%
\Omega $, and a continuous function $v$ from $%
\mathbb{R}
_{+}^{A}$ into $%
\mathbb{R}
$, such that $u(S)=v{\large (}(|S\cap C_{a}|)_{a\in A}{\large )}$ for all $%
S\in \mathcal{B}$.
\end{itemize}

The separable domain $\mathcal{S}(\Omega )$ captures the standard
microeconomic fair division model: $A$ is a set of divisible commodities,
the manna is the bundle $\omega \in 
\mathbb{R}
_{+}^{A}$ such that $\omega _{a}=|C_{a}|$ for all $a$, a share $S_{i}$ gives
to agent $i$ the amount $z_{ia}=|S_{i}\cap C_{a}|$ of commodity $a$, and the
partition $\Pi =(S_{i})_{i=1}^{n}$ corresponds to the division of the manna
as $\omega =\sum_{1}^{n}z_{i}$ .

In the general non atomic division problem, the set of shares $\mathcal{B}$
is not compact for the pseudo-metric $\delta $. It follows that when we
maximize or minimize utilities over shares, or look for a partition
achieving a benchmark utility $minMax$ or $Maxmin$, we cannot claim the
existence of an exact solution to the program: the $minMax$ is not a true
minimum, only an infimum, and $Maxmin$ is only a supremum, not a true
maximum. As this will cause no confusion, we stick to the $min$ and $Max$
notation throughout.

However in the microeconomic model, the set of shares and of partitions are
both compact so for this important set of problems (where all our examples
live) the $min$ and $Max$ notation are strictly justified.\smallskip

One can also specialise the general model by imposing constraints on the set
of feasible shares. The most important instance is the familiar\textit{\
interval model}, where the manna is $\Omega =[0,1]$ and a share must be an
interval, so an $n$-partition is made of $n$ adjacent intervals. Other
instances assume $\Omega $ is a polytope, and shares are polytopes of a
certain type: e.g. triangles or tetrahedrons (\cite{SNHA}). And sometimes
shares must be connected subsets of $\Omega $ (\cite{BDT}, \cite{AD})\textit{%
.}

The Divide and Choose$_{n}$ rules, as well as our Bid and Choose$_{n}$
rules, do not work in these models\footnote{%
For instance in the interval model, the first divider can find an
equipartition made of adjacent intervals (by our Lermma 1), but the next
agent called to divide is typically left with disconnected intervals.}, so
our Theorems 1 and 2 do not apply. But the interval model is still useful
here in a technical sense: the proof of the critical Lemma 1 in Subsection
3.2 starts by projecting the general problem onto an interval model and
proving existence of an equipartition there.

\subsection{Equipartitions}

\textbf{Definition 1 }\textit{An} $n$-\textit{equipartition of the share }$%
T\in \mathcal{B}$ \textit{for utility} $u\in \mathcal{D}(T)$ \textit{is a
partition} $\Pi ^{e}=(S_{i})_{i=1}^{n}\in \mathcal{P}_{n}(T)$ \textit{such
that }$u(S_{i})=u(S_{j})$\textit{\ for all }$i,j\in \{1,\cdots ,n\}$\textit{%
; we write }$u(\Pi ^{e})$\textit{\ for this common value, and }$\mathcal{EP}%
_{n}(T;u)$ \textit{for the set of these }$n$\textit{-equipartitions.%
\smallskip }

It is clear that $\mathcal{EP}_{n}(S;u)$ is non empty if $u$ is additive: if 
$\mathcal{B[}S\mathcal{]}$ is the subset of shares included in $S$, Lyapunov
Theorem implies that the range $u(\mathcal{B[}S\mathcal{])}$ is convex, so
it contains $\frac{1}{n}u(S)$; then we replace $n$ by $n-1$ and repeat the
argument on the remaining share.

The same is true if $u$ is monotone ($u\in \mathcal{M}^{\pm }(\Omega )$),
and the proof, outlined in Remark 1 below, is fairly simple. That of our
next statement is much harder.\smallskip

\textbf{Lemma 1 }(\cite{AvKa2})

\noindent \textit{Fix a share }$S\in \mathcal{B}$\textit{\ and a utility }$%
u\in \mathcal{D}(\Omega )$\textit{. The set }$\mathcal{EP}_{n}(S;u)$\textit{%
\ of }$n$\textit{-equipartitions of }$S$ \textit{at }$u$\textit{\ is non
empty.\smallskip }

\textbf{Proof.} The Theorem in \cite{AvKa2} proves Lemma 1 for the interval
model (which, as mentioned above, is not a special case of our model). Fix a
real valued function $f$ on the set of intervals $[a,b]\subset \lbrack 0,1]$%
, continuous in the standard topology and such that $f(a,a)=0$ for all $a\in
\lbrack 0,1]$. Then there exist $n$ subintervals $%
[0=x_{0},x_{1}],[x_{1},x_{2}],\cdots ,[x_{n-1},x_{n}=1]$ of $[0,1]$ forming
an equipartition of $f$: $f(x_{i-1},x_{i})$ is constant for $i=1,\cdots ,n$.

Start now from a share $S$ in the statement of Lemma 1 and pick a \textit{%
moving} \textit{knife} through $S$, i. e., a path $\kappa :[0,1]\ni
t\rightarrow K(t)\in \mathcal{B}$ from $K(0)=\varnothing $ to $K(1)=S$,
continuous for the pseudo-metric $\delta $ on $\mathcal{B}$ and weakly
inclusion increasing:%
\begin{equation*}
0\leq t<t^{\prime }\leq 1\Longrightarrow K(t)\subseteq K(t^{\prime })
\end{equation*}
(in Subsection 5.1 moving knifes must be strictly inclusion increasing).
Then the function%
\begin{equation*}
f(a,b)=u(K(b)\diagdown K(a))
\end{equation*}%
is as in the previous paragraph, and an $f$-equipartition $%
([x_{i-1},x_{i}])_{i=1}^{n}$ of $[0,1]$ yields the desired $u$-equipartition 
$(K(x_{i})\diagdown K(x_{i-1}))_{i=1}^{n}$ of $S$. $\blacksquare $

\textit{Remark 1 It is easy to prove Lemma 1 if we assume that the sign of }$%
u$\textit{\ is constant: all shares are weakly preferred to the empty share,
or all are weakly worse. Assume the former and use as above a moving knife
to project }$S$\textit{\ onto }$[0,1]$\textit{, where a }$n$-\textit{%
partition is identified with a point in the simplex of dimension }$n-1$.%
\textit{\ Then apply the Knaster--Kuratowski--Mazurkiewicz Lemma to the sets 
}$Q_{i}$ \textit{of partitions of the interval\ where the }$i$\textit{-th
interval gives the lowest utility: each }$Q_{i}$ \textit{is closed, contains
the }$i$\textit{-th} \textit{face of the simplex, and their union covers it
entirely. Thus these sets intersect.}

\textit{One can also invoke the stronger results in \cite{Str} and \cite{Su}
showing the existence of an Envy-free partition under this assumption. But
recall that a key feature in the division of a mixed manna is that the sign
of }$u$\textit{\ is \textbf{not} constant across shares.}

\subsection{Two utility benchmarks}

\textbf{Definition 2 }\textit{Fix }$n$\textit{, the manna }$(\Omega ,%
\mathcal{B})$ \textit{and} $u\in \mathcal{D}(\Omega )$\textit{:}%
\begin{equation}
minMax(u;n)=\min_{\Pi \in \mathcal{P}_{n}(\Omega )}\max_{1\leq i\leq
n}u(S_{i})\text{ ; }Maxmin(u;n)=\max_{\Pi \in \mathcal{P}_{n}(\Omega
)}\min_{1\leq i\leq n}u(S_{i})  \label{2}
\end{equation}%
Recall that $minMax$ is the utility agent $u$ can achieve by having first
pick in the worst possible $n$-partition of $\Omega $, and $Maxmin$ by
having last pick in the best possible $n$-partition of $\Omega $.\smallskip

\textbf{Proposition 1}

$i)$ \textit{If }$u\in \mathcal{A}dd(\Omega )$ \textit{then} $%
minMax(u;n)=Maxmin(u;n)=\frac{1}{n}u(\Omega )$

$ii)$ \textit{If} $u\in \mathcal{M}^{\pm }(\Omega )$%
\begin{equation}
minMax(u;n)=\min_{\Pi ^{e}\in \mathcal{EP}_{n}(\Omega ;u)}u(\Pi ^{e})\text{
; }Maxmin(u;n)=\max_{\Pi ^{e}\in \mathcal{EP}_{n}(\Omega ;u)}u(\Pi ^{e})
\label{12}
\end{equation}

$iii)$ \textit{If} $u\in \mathcal{D}(\Omega )$%
\begin{equation}
minMax(u;n)\leq \min_{\Pi ^{e}\in \mathcal{EP}_{n}(\Omega ;u)}u(\Pi
^{e})\leq \max_{\Pi ^{e}\in \mathcal{EP}_{n}(\Omega ;u)}u(\Pi ^{e})\leq
Maxmin(u;n)  \label{11}
\end{equation}%
\smallskip

\textbf{Proof}

\noindent \textit{Statement} $iii)$ If $\Pi ^{e}$ is an $n$-equipartition, $%
u(\Pi ^{e})$ is the utility of its best share, hence $minMax(u;n)\leq u(\Pi
^{e})$; proving the other inequality in (\ref{11}) is just as easy.

\noindent \textit{Statement }$i)$ By additivity of $u$, for any $n$%
-partition $\Pi $ we have $\max_{i}u(P_{i})\geq \frac{1}{n}u(\Omega )$
implying $minMax(u;n)\geq \frac{1}{n}u(\Omega )$; we check symmetrically $%
\frac{1}{n}u(\Omega )\geq Maxmin(u;n)$, and the conclusion follows by
comparing these inequalities to those in (\ref{11}).

\noindent \textit{Statement }$ii)$ Assume $u\in \mathcal{M}^{+}(\Omega )$;
the proof for $\mathcal{M}^{-}(\Omega )$ is identical. The continuity and
monotonicity of $u$ imply: if $S,T$ are two disjoints shares such that $%
u(S)>u(T)$, we can trim part of $S$ and add it to $T$ to get two disjoint
shares with equal utility in between $u(S)$ and $u(T)$. Expanding this
argument, if $S_{1},\cdots ,S_{k}$ and $T$ are disjoint shares such that%
\begin{equation*}
u(S_{1})=u(S_{2})=\cdots =u(S_{k})>u(T)
\end{equation*}%
we can simultaneously trim $S_{1},\cdots ,S_{k}$ keeping them of equal
utility and add the trimming to $T$, so that the resulting $k+1$ shares are
all equally good and their common utility is between the two utilities
above. Iterating this process, we see that if $\Pi =(S_{i})_{i=1}^{n}\in 
\mathcal{P}_{n}(\Omega )$\ is such that $\max_{1\leq i\leq
n}u(S_{i})>\min_{1\leq i\leq n}u(S_{j})$, we can construct an equipartition $%
\Pi ^{e}\in \mathcal{EP}_{n}(\Omega ;u)$\ such that%
\begin{equation*}
\max_{1\leq i\leq n}u(S_{i})>u(\Pi ^{e})>\min_{1\leq j\leq n}u(S_{j})
\end{equation*}%
Now fix $\varepsilon >0$, arbitrarily small, pick $\Pi =(S_{i})_{i=1}^{n}\in 
\mathcal{P}_{n}(\Omega )$ such that $\min_{1\leq j\leq n}u(S_{j})\geq
Maxmin(u;n)-\varepsilon $, and assume that $\Pi $ is not an equipartition.
By the argument above we can find $\Pi ^{e}\in \mathcal{EP}_{n}(\Omega ;u)$
such that $u(\Pi ^{e})>\min_{1\leq j\leq n}u(S_{j})$, therefore $\Pi ^{e}$
too is an $\varepsilon $-approximation of $Maxmin(u;n)$, and the right-hand
inequality in (\ref{12}) follows. The proof of the left-hand$\ $inequality
is similar. $\blacksquare \smallskip $

In the general domain $\mathcal{D}(\Omega )$, the partitions achieving the $%
Maxmin$ and $minMax$ utilities are not necessarily equipartitions. In the
microeconomic example of Section 1, Ann has single-peaked preferences and
her $minMax$ is achieved by the all-or-nothing partition $\{\varnothing
,\Omega \}$; Bob has single-dipped preferences and the same partition
delivers his $Maxmin$; but $\{\varnothing ,\Omega \}$ is not an
equipartition for either utility.\smallskip 

\textit{Remark 2:} \textit{In the interval model with a monotone utility }$u$%
\textit{, it is easy to check that any two }$n$-\textit{equipartitions have
the same utility and in turn this implies }$minMax(u;n)=Maxmin(u;n)$\textit{%
: hence this is the best Fair Guarantee.\textbf{\ }The numerical example
above can be viewed as an instance of the interval model where the two
agents are indifferent between }$[0,x]$\textit{\ and }$[1-x,1]$\textit{\ for
all }$x$\textit{: so only the inequality (\ref{11}) holds true in the
general (non monotone) interval model.}

\subsection{Fair Guarantees}

\textbf{Definition 3 }\textit{Fix the manna} $(\Omega ,\mathcal{B})$ \textit{%
and a subdomain} $\mathcal{D}^{\ast }$, $\mathcal{D}^{\ast }\subseteq 
\mathcal{D}(\Omega )$. \textit{A Fair Guarantee in} $\mathcal{D}^{\ast }$ 
\textit{is a mapping }$\Gamma :u\rightarrow \Gamma (u;n)$\textit{\ such that
for any profile} $(u_{i})_{i=1}^{n}\in (\mathcal{D}^{\ast })^{n}$ \textit{%
there exists} $\Pi =(S_{i})_{i=1}^{n}\in \mathcal{P}_{n}(\Omega )$ \textit{%
such that }$u_{i}(S_{i})\geq \Gamma (u_{i};n)$\textit{\ for all }$i$\textit{%
.\smallskip }

In Section 1 we observed, by looking at unanimity profiles, that $%
Maxmin(\cdot ;n)$ is an upper bound for \textit{any} Fair Guarantee:
inequality (\ref{14}). We also mentioned two subdomains where $Maxmin(\cdot
;n)$ itself is a (hence the optimal) Fair Guarantee: the additive domain $%
\mathcal{A}dd(\Omega )$ and the subdomain of the separable one $\mathcal{S}%
(\Omega )$ where preferences are also convex. Finally we used the Ann and
Bob microeconomic example with a single commodity to show that $Maxmin(\cdot
;n)$ is not a Fair Guarantee in $\mathcal{D}(\Omega )$, even for $n=2$ and a
one dimensional manna.

Before proving in the next Section that $minMax(\cdot ;n)$ is a Fair
Guarantee in the whole domain $\mathcal{D}(\Omega )$ we construct a
microeconomic problem with two divisible items and two agents $u_{1}$ and $%
u_{2}$ where

\begin{center}
$minMax(u_{i};2)=0<1=Maxmin(u_{i};2)$ for $i=1,2$\linebreak and $%
(minMax(u_{1}),minMax(u_{2}))$ is weakly Pareto optimal
\end{center}

\noindent This implies that for \textit{any} Fair Guarantee $\Gamma $, at
least one of $\Gamma (u_{1};2)=0$ and $\Gamma (u_{2};2)=0$ must hold. In
words, for some problems, no Fair Guarantee can reduce the gap from $minMax$
to $Maxmin$ for both agents.\footnote{%
Divide and Choose implements the utility profile $%
(minMax(u_{i};2),Maxmin(u_{j};2))$: this gap can be closed for one agent.}

The manna is $\omega =(1,1)$ and we a share as $z=(x,y)$. Both utilities are
symmetric in $x,y$: $u_{i}(x,y)=u_{i}(y,x)$ so it is enough to define them
for $x\leq y$:%
\begin{equation*}
\begin{array}{ccc}
u_{1}(z)= & 0 & \text{if }x\leq \frac{1}{2}\leq y \\ 
u_{1}(z)= & 1-2y & \text{if }x\leq y\leq \frac{1}{2} \\ 
u_{1}(z)= & 2x-1 & \text{if }\frac{1}{2}\leq x\leq y%
\end{array}%
\end{equation*}%
\begin{equation*}
\begin{array}{ccc}
u_{2}(z)= & 0 & \text{if }x\leq y\leq \frac{1}{2}\text{ or }\frac{1}{2}\leq
x\leq y \\ 
u_{2}(z)= & 2y-1 & \text{if }\frac{1}{2}\leq y\leq 1-x \\ 
u_{2}(z)= & 1-2x & \text{if }\frac{1}{2}\leq 1-x\leq y%
\end{array}%
\end{equation*}%
The range of both utilities is $[0,1]$. Agent $1$'s utility $u_{1}(z_{1})$
is null in the NW and SE quadrant of the box $[0,1]^{2}$ with center at $(%
\frac{1}{2},\frac{1}{2})$; it is strictly positive in the SW and NE
quadrants except on the lines $x=\frac{1}{2}$ and $y=\frac{1}{2}$. Agent $2$%
's utility $u_{2}(z_{2})$ is symetrically null in the SW and NE quadrants,
and strictly positive in the NW and SE quadrants except on the same two
lines. Therefore for any division $(1,1)=z_{1}+z_{2}$ of the manna we have $%
u_{1}(z_{1})\cdot u_{2}(z_{2})=0$: there is no feasible division s. t. $%
u_{i}(z_{i})>0$ for $i=1,2$.

The partition $\{(0,0),(1,1)\}$ achieves $Maxmin(u_{1})=1$ and $%
minMax(u_{2})=0$; the partition $\{(0,1),(1,0)\}$ achieves $Maxmin(u_{2})=1$
and $minMax(u_{1})=0$.

\section{The Divide \& Choose$_{n}$ rule}

Start by a combinatorial observation. Let $G$ be a bilateral graph between
the sets $M$ of agents and $R$ of shares: interpret $(m,r)\in G$ as agent $m$
\textit{likes} share $r$. We say that the subset $\widetilde{M}$ of agents
are \textit{properly matched} to the subset $\widetilde{R}$ of shares if $|%
\widetilde{M}|=|\widetilde{R}|$, agents in $\widetilde{M}$ are each matched
(one-to-one) to a share they like in $\widetilde{R}$, and no one outside $%
\widetilde{M}$ likes any share in $\widetilde{R}$.\smallskip

\textbf{Lemma 2}. \textit{Assume }$|M|=|R|$\textit{, each agent in }$M$%
\textit{\ likes at least one object in }$R$ \textit{and some agent }$i^{\ast
}$\textit{\ likes all objects in }$R$\textit{. Then there is a (non empty)
largest set }$M^{\ast }$\textit{\ of properly matchable agents containing }$%
i^{\ast }$\textit{: if }$\widetilde{M}$\textit{\ is properly matched to }$%
\widetilde{R}$\textit{, then }$\widetilde{M}\subseteq M^{\ast }$\textit{%
.\smallskip }

\textbf{Proof}. We apply the Gallai-Edmonds decomposition of a bipartite
graph: see e.g. \cite{LP} Chap 3 (or Lemma 1 in \cite{BM}). If $M$ can be
matched with $R$ this is a proper match and the statement holds true. If $M$
and $R$ cannot be matched, then we can uniquely partition $M$ as $%
(M^{+},M^{\ast })$ and $R$ as $(R^{+},R^{\ast })$ such that:

\noindent 1. $|M^{+}|>|R^{+}|$, the agents in $M^{+}$ do not like any object
in $R^{\ast }$, and they compete for the over-demanded objects in $R^{+}$:
every subset of $R^{+}$ is liked by a strictly larger subset of agents in $%
M^{+}$;

\noindent 2. $|M^{\ast }|<|R^{\ast }|$ and the agents in $M^{\ast }$ can be
matched with some subset of $R^{\ast }$.

By the general Gallai-Edmonds result, $M^{+}$ and $R^{\ast }$ are non empty.
Here $M^{\ast }$ is non empty as well because it contains the special agent $%
i^{\ast }$. Every match of $M^{\ast }$ to a subset of $R^{\ast }$ is proper.
Finally suppose $\widetilde{M}$ is properly matched to $\widetilde{R}$ and $%
\widehat{M}=\widetilde{M}\cap M^{+}$ is non empty. Then $\widehat{M}$ is
matched to some subset $\widehat{R}$ of $R^{+}$ but $\widehat{R}$ is liked
by more agents in $M^{+}$ than there are in $\widehat{M}$, therefore the
match is not proper: contradiction. So $\widetilde{M}$ does not intersect $%
M^{+}$ as was to be proved.$\blacksquare \smallskip $

\textbf{Definition 4: }\textit{the D\&C}$_{n}$\textit{\ rule}.

\noindent \textit{Fix the manna} $(\Omega ,\mathcal{B})$ \textit{and the
ordered set of agents }$N=\{1,\cdots ,n\}$\textit{, each with a utility in} $%
\mathcal{D}(\Omega )$\textit{.}

\noindent \textit{Step 1. Agent }$1$\textit{\ proposes a partition }$\Pi
^{1}\in \mathcal{P}_{n}(\Omega )$\textit{;} \textit{all other agents report
which shares in }$\Pi ^{1}$\textit{\ they like (at least one). In the
resulting bipartite graph between }$N$ \textit{and the shares in }$\Pi ^{1}$%
\textit{, where agent }$1$\textit{\ likes all the shares, we use Lemma 2 to
match properly the largest possible set of agents }$N^{1}$\textit{\ (it
contains agent }$1$\textit{) with some set of shares }$R$\textit{; if }$%
N^{1}=N$ \textit{we are done, otherwise} we\textit{\ go to}

\noindent \textit{Step 2. Repeat with the remaining manna }$\Omega ^{2}$ 
\textit{and agents in }$N\diagdown N^{1}$\textit{. Ask the first agent in
the exogenous ordering to propose a partition }$\Pi ^{2}\in \mathcal{P}%
_{n-|N_{1}|}(\Omega ^{2})$\textit{, while others report which of these new
shares they like. And so on.\smallskip }

At least one agent, the Divider, is served in each step, thus the algorithm
just described takes at most $n-1$ steps. But the algorithm matches as many
agents as possible, so as to minimize information disclosure, and typically
takes fewer steps.

There is some flexibility in the Definition of the rule: although the set of
agents matched in each step is unambiguous, we have typically several
choices for the set $R$ of shares to assign in each step, and multiple ways
to assign these to the agents.

Our first main result is that $minMax$ is a Fair Guarantee, implemented by
the D\&C$_{n}$ rule in the full domain $\mathcal{D}(\Omega )$.\smallskip

\textbf{Theorem 1}

\textit{Fix the manna} $(\Omega ,\mathcal{B})$ \textit{and} $n$.

\noindent $i)$ \textit{In the D\&C}$_{n}$ \textit{rule,} \textit{an agent
with utility }$u\in \mathcal{D}(\Omega )$ \textit{guarantees the }$%
minMax(u;n)$\textit{\ utility level by 1) when called to divide, proposing
an equipartition }$\Pi ^{e}\in \mathcal{EP}_{m}(S;u)$ \textit{of the
remaining share }$S$ \textit{of manna among the }$m$\textit{\ remaining
agents, and 2) when reporting shares he likes, accepting \textbf{all}
shares, and only those,\ not worse than }$minMax(u;n)$\textit{\ (the }$minMax
$ \textit{level in the \textbf{initial} problem}).

\noindent $ii)$ \textit{Moreover the first Divider (and no one else)
guarantees her }$Maxmin$\textit{\ utility by proposing her }$Maxmin$\textit{%
\ partition in Step 1.\textbf{\ }Other agents cannot guarantee more than
their }$minMax$ \textit{utility}.\smallskip

\textbf{Proof. }\textit{Statement}\textbf{\ }$i)$. Consider agent $u$ using
the strategy in the statement. At a step where he must report which shares
he likes among those offered at that step, he can for sure find one worth at
least $minMax(u;n)$: all shares previously assigned are worth to him
strictly less than $minMax(u;n)$, and together with the freshly cut shares
they form a partition in $\mathcal{P}_{n}(\Omega )$; in any partition at
least one share is worth $minMax(u;n)$ or more.

At a step where our agent is called to cut, he proposes to the remaining
agents an $m$-equipartition $\Pi ^{e}\in \mathcal{EP}_{m}(S;u)$ of the
remaining manna $S$. To check the inequality $u(\Pi ^{e})\geq minMax(u;n)$
note that $\Pi ^{e}$ together with the previously assigned shares is a
partition of $\Omega $ in which the old shares are worth strictly less than $%
minMax(u;n)$.\footnote{%
After Step 1 an agent can secure his $Maxmin$ utility for the smaller manna $%
S$ among $m$ agents, but this may be below the $Maxmin$ utility in the
initial problem.}\smallskip 

\textit{Statement }$ii)$. This is clear for the first Divider. Fix now an
agent $i$ with utility $u$ and check that if he is not the first Divider,
for certain moves of the other agents, agent $u$ gets exactly his $minMax$
utility. Pick a partition $\Pi \in \mathcal{P}_{n}(\Omega )$ achieving $%
minMax(u;n)$ (as usual, the existence assumption is without loss). Suppose
that the first Divider, who is not agent $i$, offers $\Pi $, and all agents
other than $i$ (including the Divider) find all shares acceptable: then a
full match is feasible ($i$ must accept at least one share) so $i$'s share
cannot be worth more than $minMax(u;n)$. $\blacksquare \smallskip $

\section{Bid and Choose and Moving Knives for good or bad manna}

We now assume that the manna is unanimously good, $u\in \mathcal{M}%
^{+}(\Omega )$, or unanimously bad, $u\in \mathcal{M}^{-}(\Omega )$. Because 
$u(\varnothing )=0$, for all $S$ we have $u(S)\geq 0$ in the former case and 
$u(S)\leq 0$ in the latter. Recall that in these two domains, the $minMax$
(resp. $Maxmin$) utility is the smallest (resp. largest) equipartition
utility: property (\ref{12}) in Proposition 1.

We check first that the profile of $Maxmin$ utility levels still may not be
feasible, even in the simple microeconomic model (corresponding to the
separable domain $\mathcal{S}(\Omega )$ in Subsection 3.1). We have one unit
each of two divisible goods,$\omega =(1,1)$, to be shared between two
agents. The first agent has Leontief preferences $u_{1}(z)=\min \{x,y\}$ so
his worst case partition is $\Pi =\{(1,0),(0,1)\}$ and his best one is the
equal split partition $\Pi ^{\prime }=\{\frac{1}{2}\omega ,\frac{1}{2}\omega
\}$: $minMax(u_{1};2)=0<\frac{1}{2}=Maxmin(u_{1};2)$. Agent $2$ has
anti-Leontief preferences:\ $u_{2}(z)=\max \{x,y\}$. For her the equal split
partition $\Pi ^{\prime }$ is the worst and the best one is $\Pi $: $%
minMax(u_{2};2)=\frac{1}{2}<1=Maxmin(u_{2};2)$. Clearly the profile of $%
Maxmin$ utilities $(\frac{1}{2},1)$ is not feasible, while D\&C$_{2}$
implements $(\frac{1}{2},\frac{1}{2})$ and $(0,1)$, depending on who is the
Divider.

We show that the $minMax$ guarantee is always improved, at least weakly, by
the large family of Bid and Choose (B\&C$_{n}$) rules, inspired by the
familiar Moving Knives (MK$_{n}$) rules (\cite{DS}).

\subsection{MK$_{n}^{\protect\kappa }$ and B\&C$_{n}^{\protect\theta }$ rules%
}

A \textit{moving} \textit{knife} through the manna $(\Omega ,\mathcal{B}%
,|\cdot |)$ is a path $\kappa :[0,1]\ni t\rightarrow K(t)\in \mathcal{B}$
from $K(0)=\varnothing $ to $K(1)=\Omega $, continuous for the pseudo-metric 
$\delta $ on $\mathcal{B}$ and strictly inclusion increasing:%
\begin{equation*}
0\leq t<t^{\prime }\leq 1\Longrightarrow K(t)\subset K(t^{\prime })\text{
and }|K(t^{\prime })\diagdown K(t)|>0
\end{equation*}%
The moving knife $\kappa $ arranges shares of increasing value to all
participants along the specific path of the knife. An example is $%
K(t)=B(t)\cap \Omega $, where $t\rightarrow B(t)$ is a path of balls with a
fixed center and radius growing from $0$ to $1$, so that $B(1)$ contains $%
\Omega $. Moving knifes can take many other shapes, for instance hyperplanes.

Our Bid and Choose rules offer more choices than Moving Knives to the
agents, with the help of a benchmark\textit{\ measure} $\theta $ of the
shares, chosen by the rule designer: $\theta $ is a positive $\sigma $%
-additive measure on $(\Omega ,\mathcal{B})$, normalised to $\theta (\Omega
)=1$. It is absolutely continuous w.r.t. the Lebesgue measure $|\cdot |$ and
vice versa: the density of $\theta $ w.r.t. $|\cdot |$ is strictly positive.
In particular $\theta $ is strictly inclusion increasing:%
\begin{equation*}
\forall S,T\in \mathcal{B}:S\subset T\text{ and }|T\diagdown S|>0\Rightarrow
\theta (S)<\theta (T)
\end{equation*}%
In applications $\theta $ can evaluate for instance the market value,
physical size, or weight of a share.

Fixing a moving knife $\kappa $ and a measure $\theta $, we define in
parallel the Moving Knife (MK$_{n}^{\kappa }$) and the Bid and Choose (B\&C$%
_{n}^{\theta }$) rules. In both cases a clock $t$ runs from $t=0$ to $t=1$%
.\smallskip

\textbf{Definition 5 }\textit{the MK}$_{n}^{\kappa }$ \textit{and\ B\&C}$%
_{n}^{\theta }$\textit{\ rules with increasing utilities}

\noindent \textit{Step 1. The first agent }$i_{1}$\textit{\ to stop the
clock, at }$t^{1}$\textit{, gets the share }$K(t^{1})$ \textit{in} \textit{MK%
}$_{n}^{\kappa }$\textit{, or in B\&C}$_{n}^{\theta }$ \textit{chooses any
share in }$\Omega $\textit{\ s.t.} $\theta (S)=t^{1}$\textit{, say} $%
S_{i_{1}}$\textit{,\ and leaves;}

\noindent \textit{Step k: Whoever stops the clock first at }$t^{k}$\textit{\
gets the share }$K(t^{k})\diagdown K(t^{k-1})$ \textit{in} \textit{MK}$%
_{n}^{\kappa }$\textit{, or in B\&C}$_{n}^{\theta }$ \textit{chooses any
share} \textit{in} $\Omega \diagdown {\large \cup }_{1}^{k-1}S_{i_{\ell }}$ 
\textit{s.t.} $\theta (S)=t^{k}-t^{k-1}$\textit{, say} $S_{i_{k}}$\textit{,} 
\textit{and leaves;}

\noindent \textit{In Step }$n-1$\textit{\ the\ single remaining agent who
did not stop the clock takes the remaining share }$\Omega \diagdown
K(t^{n-1})$ or\textit{\ }$\Omega \diagdown {\large \cup }_{1}^{n-1}S_{i_{%
\ell }}$\textit{.\smallskip }

\textbf{Definition 5}$^{\ast }$ \textit{with decreasing utilities}

\noindent \textit{In each step all agents must choose a time to
\textquotedblleft drop\textquotedblright , and the last agent }$i_{1}$%
\textit{\ who drops, at }$t^{1}$\textit{,\ gets }$K(t^{1})$\textit{\ in MK}$%
_{n}^{\kappa }$\textit{, or in B\&C}$_{n}^{\theta }$\textit{\ chooses }$%
S_{i_{1}}$\textit{\ s.t. }$\theta (S_{i_{1}})=t^{1}$\textit{. The other
steps are similarly adjusted.\smallskip }

Breaking ties between agents stopping the clock (or dropping) at the same
time is the only indeterminacy in these rules, much less severe than in D\&C$%
_{n}$, where we serve at each step an unambiguous set of agents, but there
are typically several ways to match them properly.

Up to tie-breaking, B\&C$_{n}^{\theta }$ and MK$_{n}^{\kappa }$ are
anonymous (do not discriminates between agents) but not neutral (do
discriminate between shares), while D\&C$_{n}$ is neutral but not anonymous.

In MK$_{n}^{\kappa }$ the share of an agent takes the form $K(t)\diagdown
K(t^{\prime })$ so it covers a set of dimension $2$ (and feasible partitions
move in a set of dimension $n-1$). By contrast every partition in $\mathcal{P%
}_{m}(\Omega )$ is feasible under the B\&C$_{n}^{\theta }$ rule.

To check this fix $\Pi =(S_{i})_{i=1}^{n}$ and assume first $|S_{i}|>0$ for
all $i$. Consider $n$ agents deciding (cooperatively) to achieve $\Pi $. By
the strict monotonicity of $\theta $ the sequence $t^{i}=$ $\theta ({\large %
\cup }_{j=1}^{i}S_{j})$ increases strictly therefore they can stop the clock
(or drop) at these successive times and choose the corresponding shares in $%
\Pi $. If there are shares of measure zero they can all be distributed at
time $0$.

On the other hand in B\&C$_{n}^{\theta }$ all but one agent must pick a
share under constraints, thus revealing more information than in MK$%
_{n}^{\kappa }$. Loosely speaking, B\&C$_{n}^{\theta }$ is informationally
comparable to D\&C$_{n}$.\smallskip

\textit{Remark 3. We can also implement the Guarantees described in the next
Subsection by alternative static versions of MK}$_{n}^{\kappa }$\textit{\ }%
and \textit{B\&C}$_{n}^{\theta }$\textit{\ where agents bid all at once for
potential stopping times; we do not discuss these rules for the sake of
brevity.}

\subsection{B\&C$^{\protect\theta }$ and MK$^{\protect\kappa }$ Guarantees}

We fix an increasing utility $u\in \mathcal{M}^{+}(\Omega )$. The results
are identical, and identically phrased, for a bad manna $u\in \mathcal{M}%
^{-}(\Omega )$. See also Remark 4 at the end of this Subsection.

Define the triangle $\mathcal{T}=\{(t^{1},t^{2})|0\leq t^{1}\leq t^{2}\leq
1\}$ in $%
\mathbb{R}
_{+}^{2}$ and the set $\Upsilon (n)$ of increasing sequences $\tau
=(t^{k})_{0\leq k\leq n}$ in $[0,1]$ s.t.%
\begin{equation*}
t^{0}=0\leq t^{1}\leq \cdots \leq t^{n-1}\leq 1=t^{n}
\end{equation*}%
For a moving knife $\kappa $, utilities of the shares in MK$_{n}^{\kappa }$
are described by the function $u^{\kappa }$ on $\mathcal{T}$:%
\begin{equation*}
u^{\kappa }(t^{1},t^{2})=u{\large (}K(t^{2})\diagdown K(t^{1}){\large )}%
\text{ for all }(t^{1},t^{2})\in \mathcal{T}
\end{equation*}%
For a measure $\theta $, the corresponding definition in B\&C$^{\theta }$ is
the\textit{\ indirect utility} $u^{\theta }$:%
\begin{equation}
u^{\theta }(t^{1},t^{2})=\min_{T:\theta (T)=t^{1}}\max_{S:S\cap
T=\varnothing ;\theta (S)=t^{2}-t^{1}}u(S)\text{ for all }(t^{1},t^{2})\in 
\mathcal{T}  \label{6}
\end{equation}%
Both $u^{\kappa }$ and $u^{\theta }$ decrease (weakly) in $t^{1}$ and
increase (weakly) in $t^{2}$.

We show below that the Guarantees $\Gamma ^{k}$ and $\Gamma ^{\theta }$
implemented by MK$_{n}^{\kappa }$ and B\&C$_{n}^{\theta }$ respectively are
computed as%
\begin{equation}
\Gamma ^{\alpha }(u;n)=\max_{\tau \in \Upsilon (n)}\min_{0\leq k\leq
n-1}u^{\alpha }(t^{k};t^{k+1})\text{ where }\alpha \text{ is }\kappa \text{
or }\theta  \label{5}
\end{equation}

For instance in MK$_{2}^{\kappa }$ with two agents, write $\tau _{\kappa }$
for the (not necessarily unique) position of the knife making our agent
indifferent between the share $K(\tau _{\kappa })$ and its complement. Then%
\begin{equation*}
\Gamma ^{\kappa }(u;2)=\max_{0\leq t^{1}\leq 1}\min \{u(K(t^{1})),u(\Omega
\diagdown K(t^{1}))=u(K(\tau _{\kappa }))=u(\Omega \diagdown K(\tau _{\kappa
})
\end{equation*}%
In B\&C$_{2}^{\theta }$ the bid $\tau _{\theta }$ makes the best share of
size $\tau _{\theta }$ as good as the worst share of size $1-\tau _{\theta }$%
:%
\begin{equation}
\Gamma ^{\theta }(u;2)=\max_{0\leq t^{1}\leq 1}\min \{\max_{\theta
(S)=t^{1}}u(S),\min_{\theta (S)=t^{1}}u(\Omega \diagdown S)\}=\max_{\theta
(S)=\tau _{\theta }}u(S)=\min_{\theta (S)=\tau _{\theta }}u(\Omega \diagdown
S)  \label{9}
\end{equation}

\textbf{Lemma 4}

\noindent $i)$\textit{\ The utility }$u^{\kappa }$\textit{\ and the indirect
utility }$u^{\theta }$ \textit{are continuous. Both the minimum and maximum
in (\ref{6}) are achieved.}

\noindent $ii)$ \textit{The maximum of problem (\ref{5}) (for both rules) is
achieved at some }$\tau \in \Upsilon (n)$\textit{\ where the sequence }$%
t^{k} $ \textit{increases in }$k$, \textit{all the }$u^{\alpha
}(t^{k};t^{k+1})$\textit{\ are equal, and this common utility is the optimal
value of (\ref{5}).}

Proof in the Appendix.\smallskip 

\textbf{Theorem 2 }

\noindent \textit{Fix the manna }$(\Omega ,\mathcal{B})$\textit{, the number
of agents }$n$\textit{, and a utility }$u\in \mathcal{M}^{+}(\Omega )$.

\noindent $i)$ \textit{With the MK}$_{n}^{\kappa }$\textit{\ rule, an agent
guarantees the utility }$\Gamma ^{\kappa }(u;n)$\textit{\ by committing to
stop the knife at }$t_{\kappa }^{k}$\textit{\ if exactly }$k-1$\textit{\
other agents have been served before;}

\noindent $ii)$ \textit{With the B\&C}$_{n}^{\theta }$\textit{\ rule, she
guarantees }$\Gamma ^{\theta }(u;n)$\textit{\ by stopping the clock at }$%
t_{\theta }^{k}$\textit{\ if exactly }$k-1$\ \textit{other agents have been
served before; and choosing then the best available share of size }$%
t^{k}-t^{k-1}$\textit{.}

\noindent $iii)$ $minMax(u;n)\leq \Gamma ^{\alpha }(u;n)\leq Maxmin(u;n)$
where $\alpha $ is $\kappa $ or $\theta $.

\textbf{Proof}.

\noindent \textit{Statement }$i)$ \textit{and }$iii)$\textit{\ for MK}$%
_{n}^{\kappa }$. Recall the equipartition $\Pi ={\large (}K(t_{\kappa
}^{k})\diagdown K(t_{\kappa }^{k-1}){\large )}_{1}^{n}$ has $u(\Pi )=\Gamma
^{\kappa }(u;n)$. Thus (\ref{12}) in Proposition 1 implies the inequalities $%
iii)$. Next if the knife has been stopped $k-1$\textit{\ }times before our
agent is served, the last stop occured at or before $t_{\kappa }^{k-1}$
therefore if she does stop the knife at $t_{\kappa }^{k}$ (and wins the
possible tie break) her share is at least $K(t_{\kappa }^{k})\diagdown
K(t_{\kappa }^{k-1})$. If she never gets to stop the knife, the last stop is
at or before $t_{\kappa }^{n-1}$ and she gets at least $\Omega \diagdown
K(t_{\kappa }^{n-1})$.\textit{\smallskip }

\noindent \textit{Statement} $ii)$. If she is the first to stop the clock
(perhaps also winning the tie break) at step $k$, in step $k-1$ the clock
stopped at $t^{k-1}\leq t_{\theta }^{k-1}$ and the share $T$ already
distributed at that time has $\theta (T)=t^{k-1}$: therefore she can choose
a share with utility $u^{\theta }(t^{k-1};t_{\theta }^{k})\geq u^{\theta
}(t_{\theta }^{k-1};t_{\theta }^{k})=\Gamma ^{\theta }(u;n)$. If she is the
last to be served, having never stopped the clock (or lost some tie breaks)
the share assigned to all other agents has $\theta (T)=t^{n-1}\leq t_{\theta
}^{n-1}$ therefore her share is worth $u^{\theta }(t^{n-1};1)\geq u^{\theta
}(t_{\theta }^{n-1};1)=\Gamma ^{\theta }(u;n)$.\smallskip

\noindent \textit{Statement }$iii)$ \textit{for B\&C}$_{n}^{\theta }$.

\textit{Right hand inequality}. It is enough to construct a partition $\Pi
=(S_{k})_{1}^{n}$ in which the utility of every share $S_{k},0\leq k\leq n-1$
is at least $\ u^{\theta }(t_{\theta }^{k-1},t_{\theta }^{k})$, implying $%
\min_{k}u(S_{k})\geq \Gamma ^{\theta }(u;n)$. We proceed by induction on the
steps of B\&C$_{n}^{\theta }$. First $S_{1}$ maximizes $u(S)$ s.t. $\theta
(S)=t_{\theta }^{1}$ so $u(S_{1})=u^{\theta }(0;t_{\theta }^{1})=\Gamma
^{\theta }(u;n)$ and $\theta (S_{1})=t_{\theta }^{1}$. Assume the sets $%
S_{\ell }$ are constructed for $1\leq \ell \leq k$, mutually disjoint, s.t. $%
\theta (S_{\ell })=t_{\theta }^{\ell }-t_{\theta }^{\ell -1}$ and $u(S_{\ell
})\geq u^{\theta }(t_{\theta }^{\ell },t_{\theta }^{\ell -1})$: then the set 
$T={\large \cup }_{1}^{k}S_{\ell }$ is of size $t_{\theta }^{k}$ and we pick 
$S_{k+1}$ maximizing $u(S)$ s.t. $S\cap T=\varnothing $ and $\theta
(S)=t_{\theta }^{k+1}-t_{\theta }^{k}$. By definition (\ref{6}) we have $%
u(S_{k})\geq u^{\theta }(t_{\theta }^{k};t_{\theta }^{k+1})$ and the
induction proceeds. Note that in fact $\min_{k}u(S_{k})=\Gamma ^{\theta
}(u;n)$.\smallskip

\textit{Left hand inequality}. We need now construct a partition $\Pi
=(R_{k})_{1}^{n}$ s. t. $u(R_{k})\leq u^{\theta }(t_{\theta
}^{k-1};t_{\theta }^{k})$ for $1\leq k\leq n$. We do this by a decreasing
induction in $n$. In (the first) step $n$ of the induction we define the $2$%
-partition $\Pi ^{n}=(T_{n-1},R_{n})$ of $\Omega $ where $T_{n-1}$ is any
solution of the program $\min_{T:\theta (T)=t_{\theta }^{n-1}}u(\Omega
\diagdown T)$, and $R_{n}=\Omega \diagdown T_{n-1}$. Thus $%
u(R_{n})=u^{\theta }(t_{\theta }^{n-1};1)$ and $\theta (T_{n-1})=t_{\theta
}^{n-1}$.

Assume that in step $k$ we constructed the $(n-k+2)$-partition $\Pi
^{k}=(T_{k-1},R_{k},R_{k+1},\cdots ,R_{n})$ s.t. $\theta (T_{k-1})=t_{\theta
}^{k-1}$ and $u(R_{\ell })\leq u^{\theta }(t_{\theta }^{\ell -1};t_{\theta
}^{\ell })$ for $k\leq \ell \leq n$. Pick $\widetilde{T}$ a solution of%
\begin{equation*}
\min_{T:\theta (T)=t_{\theta }^{k-2}}\max_{S:S\cap T=\varnothing ;\theta
(S)=t_{\theta }^{k-1}-t_{\theta }^{k-2}}u(S)=u^{\theta }(t_{\theta
}^{k-2};t_{\theta }^{k-1})
\end{equation*}%
As $\theta (\widetilde{T}\cap T_{k-1})\leq t_{\theta }^{k-2}$ and $\theta
(T_{k-1})=t_{\theta }^{k-1}$ we can choose $T_{k-2}$ s.t. $\widetilde{T}\cap
T_{k-1}\subseteq T_{k-2}\subseteq T_{k-1}$ and $\theta (T_{k-2})=t_{\theta
}^{k-2}$. Then we set $R_{k-1}=T_{k-1}\diagdown T_{k-2}$ so that $%
u(R_{k-1})\leq u^{\theta }(t_{\theta }^{k-2};t_{\theta }^{k-1})$ follows
from $R_{k-1}\cap \widetilde{T}=\varnothing $ and the definition of $%
\widetilde{T}$. This completes the induction step. We note finally that each
set $R^{k}$ thus constructed is of $\theta $-size $t_{\theta }^{k}-t_{\theta
}^{k-1}$, and that $\max_{k}u(S_{k})=\Gamma ^{\theta }(u;n)$. $\blacksquare $%
\smallskip

It is easy to check that no agent can secure more utility than $\Gamma
_{n}^{\kappa }$ in MK$_{n}^{\kappa }$ or $\Gamma _{n}^{\theta }$ in B\&C$%
_{n}^{\theta }$.\smallskip

\textit{Remark 4. The }$minMax$\textit{\ Guarantee and }$Maxmin$\textit{\
upper bound for} $u\in \mathcal{M}^{\varepsilon }(\Omega )$ \textit{and }$%
-u\in \mathcal{M}^{-\varepsilon }(\Omega )$\textit{, where }$\varepsilon
=\pm $\textit{, are related: }$minMax(-u;n)=-Maxmin(u;n)$\textit{. With two
agents the Guarantees }$\Gamma ^{\kappa }(u;2)$\textit{\ and }$\Gamma
^{\theta }(u;2)$\textit{\ are similarly antisymmetric:}%
\begin{equation}
\Gamma ^{\alpha }(-u;2)=-\Gamma ^{\alpha }(u;2)\text{ where }\alpha \text{
is }\kappa \text{ or }\theta  \label{15}
\end{equation}%
\textit{This is clear for }$\Gamma ^{\kappa }$\textit{\ and we check it for }%
$\Gamma ^{\theta }$\textit{\ by means of the change of variable }$%
S\rightarrow S^{\prime }=\Omega \diagdown S$\textit{:}%
\begin{equation*}
\Gamma ^{\theta }(-u;2)=-\min_{0\leq t^{1}\leq 1}\max \{\min_{\theta
(S)=t^{1}}u(S),\max_{\theta (S)=t^{1}}u(\Omega \diagdown S)\}=
\end{equation*}%
\begin{equation*}
-\min_{0\leq t^{1}\leq 1}\max \{\min_{\theta (S^{\prime })=1-t^{1}}u(\Omega
\diagdown S^{\prime }),\max_{\theta (S^{\prime })=1-t^{1}}u(S^{\prime })\}
\end{equation*}%
\begin{equation*}
=-\min_{0\leq t^{\prime }\leq 1}\max \{\max_{\theta (S^{\prime })=t^{\prime
}}u(S^{\prime }),\min_{\theta (S^{\prime })=t^{\prime }}u(\Omega \diagdown
S^{\prime })\}
\end{equation*}%
\textit{and the claim follows because if two continuous functions }$%
t\rightarrow f(t)$\textit{\ and }$t\rightarrow g(t)$\textit{\ intersect in }$%
[0,1]$\textit{\ and one increases while the other decreases, then }$%
\min_{0\leq t\leq 1}\max \{f(t),g(t)\}=\max_{0\leq t\leq 1}\min \{f(t),g(t\}$%
\textit{.}

\textit{The identity (\ref{15}) generalises to }$n\geq 3$\textit{\ for the MK%
}$^{\kappa }$ \textit{Guarantee, but not for the B\&C}$^{\theta }$ \textit{%
one.}

\subsection{Microeconomic fair division}

We must divide a good manna $\omega \in 
\mathbb{R}
_{+}^{K}$ in $n$ shares $z_{i}\in 
\mathbb{R}
_{+}^{K}$. Utilities $u\in \mathcal{M}^{+}(\omega )$ are continuous and
weakly increasing on $[0,\omega ]$.

A Moving Knife is a continuous increasing path $t\rightarrow K(t)$ from $0$
to $\omega $: a natural choice is $K(t)=t\omega ,0\leq t\leq 1$: the
corresponding Guarantee $\Gamma ^{\kappa }(u;n)=u(\frac{1}{n}\omega )$ is
the \textit{Equal Split }utility $\Gamma ^{es}(u;n)=u(\frac{1}{n}\omega )$.
A positive, additive measure $\theta $ defining B\&C$^{\theta }$ is a
\textquotedblleft price\textquotedblright\ $\theta (z)=p\cdot z$, $p\in 
\mathbb{R}
_{+}^{K}\diagdown \{0\}$, so we write the corresponding Guarantee as $\Gamma
^{p}$\textbf{.}

Recall from Section 1 that if an agent's preferences are convex her Equal
Split\ utility equals her $Maxmin$ utility, the upper bound on all Fair
Guarantees ((\ref{14})), in particular it is weakly larger than the B\&C$%
^{p} $ guarantee for any $p$. The converse inequality holds for
\textquotedblleft concave preferences\textquotedblright .\smallskip

\textbf{Lemma 5}

\noindent $i)$ \textit{If the upper contours of the utility} $u\in \mathcal{M%
}^{+}(\omega )$\textit{\ are convex, then} $\Gamma ^{p}(u;n)\leq u(\frac{1}{n%
}\omega )=Maxmin(u;n)$.

\noindent $ii)$ \textit{If the lower contours of the utility} $u\in \mathcal{%
M}^{+}(\omega )$ \textit{are convex, then} $minMax(u;n)=u(\frac{1}{n}\omega
)\leq \Gamma ^{p}(u;n)$.\smallskip

The equality in statement $i)$ was proven in Section 1. A symmetrical
argument gives statement $ii)$.\smallskip

We turn to a handful of numerical examples where $K=2$, $\omega =(1,1)$, and 
$p\cdot z=\frac{1}{2}(x+y)$. Shares are $z=(x,y)$, utilities are $1$%
-homogenous and normalised so that $u(\omega )=10$. We compute our three
Guarantees: Bid and Choose $\Gamma ^{p}$, Equal Split, and $minMax$, and
compare them to the $Maxmin$ upper bound.

The first three utilities (Leontief, Cobb Douglas and CES) define convex
preferences, the last two define \textquotedblleft concave
preferences\textquotedblright\ (represented by quadratic and
\textquotedblleft anti-Leontief\textquotedblright\ utilities).

Our first table assumes two agents, $n=2$, and illustrates Lemma 5. An agent
with convex (resp. concave) preferences gets a better Guarantee under Equal
Split (resp. Bid and Choose):%
\begin{equation*}
\begin{array}{ccccc}
u(x,y) & minMax(u;2) & \Gamma ^{p}(u;2) & u(\frac{1}{2}\omega ) & Maxmin(u;2)
\\ 
10\min \{x,y\} & 0 & 3.3 & 5 & 5 \\ 
10\sqrt{x\cdot y} & 0 & 4.1 & 5 & 5 \\ 
\frac{5}{2}(\sqrt{x}+\sqrt{y})^{2} & 2.5 & 4.4 & 5 & 5 \\ 
5(x+y) & 5 & 5 & 5 & 5 \\ 
5\sqrt{2(x^{2}+y^{2})} & 5 & 5.9 & 5 & 7.1 \\ 
10\max \{x,y\} & 5 & 6.7 & 5 & 10%
\end{array}%
\end{equation*}

The equal split partition delivers the $Maxmin$ utility for the first four
preferences, and the $minMax$ utilities for the last three. The
equipartition $\Pi =\{(1,0),(0,1)\}$ gives similarly the $minMax$ utilities
of the first four, and the $Maxmin$ ones for the last three.

To compute $\Gamma ^{p}(u;2)$ we know from (\ref{9}) that the optimal bid $%
t^{1}$ (denoted $t$ for simplicity) solves%
\begin{equation*}
\max_{\frac{1}{2}(x+y)\leq t}u(x,y)=\min_{\frac{1}{2}(x+y)\leq
t}u(1-x,1-y)=\min_{\frac{1}{2}(x+y)\geq 1-t}u(x,y)
\end{equation*}%
This equality implies $0\leq t\leq \frac{1}{2}$. If $u$ represents convex
preferences symmetric in the two goods, $u(x,y)$ is maximal under $\frac{1}{2%
}(x+y)\leq t$ at $x=y=t$, and minimal under $x+y\geq 2(1-t)$ at $x=1,y=1-2t$%
. So we must solve $u(t,t)=u(1,1-2t)$: \textit{see Figure 2}.

If $u$ represents concave symmetric preferences its maximum under $\frac{1}{2%
}(x+y)\leq t$ is at $x=0,y=2t$, and its minimum under $x+y\geq 2(1-t)$ at $%
x=y=1-t$, so we solve $u(0,2t)=u(1-t,1-t)$: \textit{see Figure 3}.

We compute finally the same Guarantees with three agents:%
\begin{equation*}
\begin{array}{ccccc}
u(x,y) & minMax(u;3) & \Gamma ^{p}(u;3) & u(\frac{1}{3}\omega ) & Maxmin(u;3)
\\ 
10\min \{x,y\} & 0 & 2 & 3.3 & 3.3 \\ 
10\sqrt{x\cdot y} & 0 & 2.4 & 3.3 & 3.3 \\ 
\frac{5}{2}(\sqrt{x}+\sqrt{y})^{2} & 2 & 2.5 & 3.3 & 3.3 \\ 
5(x+y) & 3.3 & 3.3 & 3.3 & 3.3 \\ 
5\sqrt{2(x^{2}+y^{2})} & 3.3 & 4.1 & 3.3 & 4.1 \\ 
10\max \{x,y\} & 3.3 & 5 & 3.3 & 5%
\end{array}%
\end{equation*}

The $minMax$ equipartition for $u=\frac{5}{2}(\sqrt{x}+\sqrt{y})^{2}$ and
the $Maxmin$ equipartition for $u^{\prime }=5\sqrt{2(x^{2}+y^{2})}$ have the
same form $\Pi =\{(x,0),(0,x),(1-x,1-x)\}$: in the former case we find $x=%
\frac{4}{5}$ and $minMax(u;3)=2$, in the latter we get $x=2-\sqrt{2}$ and $%
Maxmin(u^{\prime };3)=10(\sqrt{2}-1)$. Lemma 5 and the partition $\Pi
^{\prime }=\{(1,0),(0,\frac{1}{2}),(0,\frac{1}{2})\}$ fill the remaining
values of $minMax$ and $Maxmin$.

To compute $\Gamma ^{p}(u;3)$\ we know by Lemma 4 that the three terms in (%
\textit{\ref{5}}) are equal. They are

$u^{p}(0,t^{1})=\max_{\frac{1}{2}(x+y)\leq t^{1}}u(x,y)$

$u^{p}(t^{1},t^{2})=\min_{\frac{1}{2}(x^{1}+y^{1})\leq t^{1}}\max_{\frac{1}{2%
}(x+y)\leq t^{2}-t^{1}\text{ and }(x^{1}+x,y^{1}+y)\leq (1,1)}u(x,y)$

$u^{p}(t^{2},1)=\min_{\frac{1}{2}(x^{2}+y^{2})\leq t^{2}}u(1-x^{2},1-y^{2})$

Clearly $t^{1}\leq \frac{1}{3}$ (as $t^{2}-t^{1}<\frac{1}{3}<t^{1}$ and $%
1-t^{2}<\frac{1}{3}<t^{1}$ are both impossible). Therefore $%
u^{p}(0,t^{1})=u^{p}(t^{1},t^{2})$ is achieved by $t^{2}=2t^{1}$ (the
constraint $(x^{1}+x,y^{1}+y)\leq (1,1)$ does not bind). Writing $%
t=t^{1}=t^{2}-t^{1}$ it remains to solve%
\begin{equation*}
\max_{\frac{1}{2}(x+y)\leq t}u(x,y)=\min_{\frac{1}{2}(x^{2}+y^{2})\leq
2t}u(1-x^{2},1-y^{2})=\min_{\frac{1}{2}(x+y)\geq 1-2t}u(x,y)
\end{equation*}

When $u$ represents convex preferences symmetric in the two goods, the
minimum on the right-hand side is achieved by $(x,y)=(1-4t,1)$ so we solve $%
u(t,t)=u(1-4t,1)$. \textit{See Figure 4.}

If $u$ represents concave symmetric preferences, the minimum on the
right-hand side is achieved by $(x,y)=(1-2t,1-2t)$ so we solve $%
u(2t,0)=u(1-2t,1-2t)$. \textit{See Figure 5}.

\section{Concluding comments}

\paragraph{Comparing B\&C$_{n}$ versus D\&C$_{n}$ rules}

The exogenous ordering of the agents greatly affects the outcome of D\&C$%
_{n} $, whereas B\&C$_{n}$ treats the agents symmetrically. On the other
hand the choice of the benchmark measure in B\&C$_{n}$ is exogenous, which
allows much, perhaps too much flexibility to the designer.

In D\&C$_{n}$ the dividing agent may have many different strategies
guaranteeing her $minMax$ utility. By contrast in B\&C$_{n}$ the solution to
programs (\ref{9}) and (\ref{5}) is often unique. Multiple choices and the
resulting indeterminacy of the outcome may be appealing for the sake of
privacy preservation, less so from the implementation viewpoint.

\paragraph{Two challenging open questions}

1). Fix the manna $(\Omega ,\mathcal{B})$ as in Theorem 1, and each of the $n
$ agents with his own utility in $\mathcal{D}(\Omega )$. As mentioned in
Section 2 and Subsection 3.2, Stromquist (\cite{Str}) showed that an
Envy-free partition of $\Omega $ exists \textit{if all utilities are non
negative for all shares}. Without the sign assumption on utilities,
Avvakumov and Karasev (\cite{AvKa}) prove existence of an Envy-free
partition if $n$ is a power of a prime number. Whether this remains true for
all $n$ is still an open question.\smallskip 

2) If the utilities vary in a domain $\mathcal{U(}\Omega \mathcal{)}$ where
the $Maxmin$ utility is not feasible, we would like to describe the family
of \textit{undominated} Fair Guarantees $u\rightarrow \Gamma (u;n)$. For
instance in the microeconomic domain $\mathcal{M}^{+}(\omega )$ of
Subsection 5.3, the Equal Split Guarantee is clearly undominated. We
conjecture that in the domains $\mathcal{M}^{\pm }(\Omega )$ the B\&C
Guarantees $\Gamma ^{\theta }$ (Subsection 5.2) are undominated as well.

\section{Appendix: proof of Lemma 4}

1). \textit{First statement.} Recall that we can replace in definition (%
\textit{\ref{6}}) the equalities like $\theta (T)=t^{1}$ with inequalities $%
\theta (T)\leq t^{1}$. We check first that the correspondence $t\rightarrow
\{S\in \mathcal{B}|\theta (S)\leq t\}$ is continuous. Upper hemi continuity
follows by the continuity of $\theta $. For lower hemi continuity pick a
sequence $t_{n}$ converging to $t$ and $S\in \mathcal{B}$ s.t. $\theta
(S)\leq t$. If $t_{n}$ has a decreasing subsequence, we set $S_{n}=S$ so
that $\theta (S_{n})\leq t_{n}$ and $S_{n}$ converges to $S$. If $t_{n}$ has
an increasing subsequence we construct an inclusion increasing sequence $%
S_{m}$ converging to $S$ and s.t. $|S_{m}|<|S|$ for all $m$: because $\theta 
$ increases strictly, so does the sequence $\theta (S_{m})$ converging to $%
\theta (S)$, therefore we can pick subsequences $S_{p}$ of $S_{m}$ and $%
t_{p} $ of $t_{n}$ s.t. $\theta (S_{p})\leq t_{p}$, as desired.

Next we apply the Maximum Theorem twice. The first one to show that the
function $(T,t^{1},t^{2})\rightarrow C(T,t^{1},t^{2})=\max \{u(S)|S\subset
\Omega \diagdown T;\theta (T\cup S)\leq t^{1}+t^{2}\}$ is continuous because
the correspondence $(T,t^{1},t^{2})\rightarrow \{S|S\subset \Omega \diagdown
T;\theta (T\cup S)\leq t^{1}+t^{2}\}$ is continuous. The second one to
deduce that the function $\min_{T:\theta (T)\leq t^{1}}C(T,t^{1},t^{2})$ is
continuous.\smallskip

2). \textit{Second statement. }For simplicity we assume $n=3$, the general
proof is entirely similar. Fixing $u$ and $t^{1}$ there is some $t^{2}$ such
that $u^{\theta }(t^{1};t^{2})=u^{\theta }(t^{2};1)$. This is because of the
monotonicity properties of $u^{\theta }$ and of the inequalities $u^{\theta
}(t^{1};t^{1})=0\leq u^{\theta }(t^{1};1)$ and $u^{\theta }(t^{1};1)\geq
0=u^{\theta }(1;1)$. This common value is unique (though $t^{2}$ may not be)
and defines a function $g(t^{1})=u^{\theta }(t^{1};t^{2})=u^{\theta
}(t^{2};1)$. It is easy to check from the continuity and monotonicity
properties of $u^{\theta }$ that $g$ is weakly decreasing and continuous.
Then we find in the same way $t^{1}$ s.t. $g(t^{1})=u^{\theta }(0;t^{1})$.

Check finally that if $\tau _{\ast }\in \Upsilon (n)$ is such that all terms 
$u^{\theta }(t_{\ast }^{k};t_{\ast }^{k+1})$, $0\leq k\leq n-1$, equal a
common value $\lambda $, then $\tau _{\ast }$ solves program (\ref{5}). If
it does not there is a $\tau $ such that $u^{\theta }(t^{k};t^{l+1})>\lambda 
$ for $0\leq k\leq n-1$. Applying this inequality at $k=0$ gives $%
t^{1}>t_{\ast }^{1}$; next at $k=1$ we get $u^{\theta
}(t^{1},t^{2})>u^{\theta }(t_{\ast }^{1},t_{\ast }^{2})$ implying $%
t^{2}>t_{\ast }^{2}$; and so on until we reach a contradiction with the fact
that both $\tau $ and $\tau _{\ast }$ are in $\Upsilon (n)$.

Finally, the optimal sequence $t^{k}$ increases in $k$, strictly if $u$ is
not everywhere zero because $u(t,t)=0$ for all $t$.

\end{document}